\documentclass[%
amsmath,amsfonts,amssymb,
aps,prx,twocolumn,longbibliography
]{revtex4-2}

\usepackage{graphicx}
\usepackage{dcolumn}
\usepackage{bm}
\usepackage{dsfont}
\usepackage{physics} 
\usepackage{upgreek} 
\usepackage[colorlinks=true,allcolors=blue]{hyperref}
\usepackage{xcolor}
\usepackage{url}
\usepackage{multirow}
\usepackage{hhline}

\newcommand{\centeredhrulefill}{\leavevmode\leaders\hrule height 0.7ex depth \dimexpr0.4pt-0.7ex\hfill\kern0pt}

\newcommand{\argmax}[1]{\text{arg\,max}\left\lbrace{#1}\right\rbrace}

\newcommand{\dL}{\text{L}}
\newcommand{\dT}{\text{T}}
\newcommand{\dA}{\text{A}}
\newcommand{\dB}{\text{B}}
\newcommand{\dR}{\text{R}}
\newcommand{\dML}{\text{ML}}
\newcommand{\dQOC}{\text{QOC}}
\newcommand{\dsep}{\text{sep}}
\newcommand{\dmax}{\text{max}}

\newcommand{\deff}{\text{eff}}

\newcommand{\dth}{\text{th}}
\newcommand{\drec}{\text{rec}}
\newcommand{\dc}{\text{c}}
\newcommand{\dexp}{\text{exp}}

\newcommand{\hatrho}{\hat{\rho}}
\newcommand{\hatE}{\hat{E}}

\newcommand{\hatU}{\hat{U}}

\newcommand{\hatx}{\hat{x}}

\newcommand{\hatp}{\hat{p}}

\begin{document}

\title{Phase-space distributions of Bose-Einstein condensates in an optical lattice: \\ Optimal shaping and reconstruction}

\author{N. Dupont$^1$, F. Arrouas$^1$, L. Gabardos$^1$, N. Ombredane$^1$, J. Billy$^1$, B. Peaudecerf$^1$, D. Sugny$^2$, D. Gu\'ery-Odelin$^1$}

\email{dgo@irsamc.ups-tlse.fr}

\affiliation{
$^1$ Laboratoire Collisions Agr\'egats R\'eactivit\'e, UMR 5589, FERMI, UT3, Universit\'e de Toulouse, CNRS,\\
118 Route de Narbonne, 31062 Toulouse CEDEX 09, France \\
$^2$ Laboratoire Interdisciplinaire Carnot de Bourgogne, UMR 6303,\\
9 Avenue A. Savary, BP 47 870, F-21078 Dijon Cedex, France
}

\date{\today}

\begin{abstract}
We apply quantum optimal control to shape the phase-space distribution of Bose-Einstein condensates in a one-dimensional optical lattice.
By a time-dependent modulation of the lattice position, determined from optimal control theory, we prepare, in the phase space of each lattice site, translated and squeezed Gaussian states, and superpositions of Gaussian states.
Complete reconstruction of these non-trivial states is performed through a maximum likelihood state tomography.
As a practical application of our method to quantum simulations, we initialize the atomic wavefunction in an optimal Floquet-state superposition to enhance dynamical tunneling signals.
\end{abstract}

\maketitle

\section{Introduction}

In current endeavors to harness quantum properties for enhanced metrology or quantum simulations, a shared requirement is the ability to prepare, manipulate, and perform measurements on complex quantum states~\cite{Cirac2012}.
Regarding the initial preparation stage, a key example in quantum metrology is given by the use of squeezed states, that allow to reach sensitivities below the standard quantum limit. This is famously the case of squeezed light, used to enhanced spectroscopy~\cite{polzik_1992} and interferometry~\cite{tse_2019_short,acernese_2019_short}, but there is also a long going theoretical and experimental effort to harness squeezing with matter  -- with effective spins derived from internal states or from states of motion, as well as with matter waves -- for enhanced matter wave interferometry~\cite{kitagawa_1993,wineland_1994,sinatra_2022,esteve_2008, lucke_2011, Kovachy2015}.
Likewise, quantum simulations, especially when relying on engineered effective Hamiltonians~\cite{dalibard_2011,goldman_dalibard_2014}, or exploiting synthetic dimensions~\cite{Ozawa2019}, benefit from the ability to prepare specific initial states of the effective system, for which adiabatic preparation methods may not exist.

Bose-Einstein condensates (BECs) constitute a platform particularly well-suited to quantum simulations as well as quantum metrology, thanks to their high level of controlability.
We focus here on engineering the motional state of BECs in a one-dimensional optical lattice using quantum optimal control (QOC) (see~\cite{boscain_2021,koch_2022} and references therein) in order to produce states with various phase-space distributions. Such optimal control processes were already implemented experimentally with success in~\cite{dupont_2021} for the control of populations and phases of momentum superpositions. We point out that similar approaches were used in a series of papers for quantum interferometry~\cite{saywell2020,frank2014,weidner2018} or quantum simulation purposes~\cite{zhou2018,frank2016}. In this study, we show that the techniques proposed in~\cite{dupont_2021} can also be applied to generate complex quantum states corresponding to a specific phase-space distribution. With our QOC protocol, and manipulating only external degrees of freedom, we are able to prepare states with exotic density and momentum distributions, which could not be reached through standard adiabatic methods.

Since the produced states cannot be easily identified from a few measurements (as in~\cite{dupont_2021}), they therefore require a full experimental characterization in order to verify the quality of the preparation.
Such a quantum state tomography~\cite{acharya_2019} is a matter of great relevance to quantum computation and simulation. Several methods to solve that problem have been put forward, using, for example, mappings from the motional state to internal degrees of freedom~\cite{leibfried_1996,fluhmann_2020,winkelmann_2022}, or, more recently, in the context of many-body systems, exploiting randomized measurements~\cite{kokail_2021} or neural networks~\cite{torlai_2018}.
Here we implement state reconstruction for the atomic state in the lattice through a maximum likelihood iterative method inspired by quantum optics~\cite{lvovsky_2001,rehacek_2001,lvovsky_2004,deleglise_2008,Evrard2021,brown_2022}, using measurements of the free evolution of the prepared state in the lattice. We demonstrate the efficiency of this approach which is original in this context and well adapted to our experimental setup. Finally, an application of this generic preparation procedure in quantum simulation is proposed in which the wave function is brought into an optimal Floquet-state superposition to enhance dynamical tunneling signals.

The paper is organized as follows. In section \ref{sec:setup_and_algo} we present our experimental setup as well as the methods employed to prepare and reconstruct quantum states. In sections \ref{sec:unsqueezed_states} and \ref{sec:squeezing} we tailor the distributions of BECs in the $(x,p)$ phase space of a one dimensional lattice, performing translation, squeezing and superposition of Gaussian states. Finally, in section \ref{sec:dynamical_tunneling} we apply our method to the preparation of an optimal initial state  for the observation of dynamical tunneling in a modulated optical lattice~\cite{heller_1981, arnal_2020}.

\section{Experimental setup and algorithms} \label{sec:setup_and_algo}

\subsection{Experimental setup}

The experiment starts with a $^{87}$Rb BEC of $5\cdot10^5$ atoms obtained in a hybrid trap formed by a crossed optical dipole trap and a magnetic quadrupole trap~\cite{fortun_2016}. The BEC is adiabatically loaded in a far-detuned one-dimensional optical lattice of period $d$ produced by two counterpropagating laser beams of wavelength $\lambda = 2d = 1064$ nm. Along the axis of the optical lattice, the atoms experience the potential
\begin{align}
V(x,t) &= -\frac{s}{2}E_\dL \cos\left( k_\dL x + \varphi(t) \right) + V_\text{hyb}(x),\label{eq:global_potential}\\ 
&=V_\dL(x,t)+ V_\text{hyb}(x),\nonumber
\end{align}
\noindent where $k_\dL = 2\pi/d$ and $E_\dL = \hbar^2 k_\dL^2/2m$ are respectively the wavenumber and characteristic energy scale of the lattice (with $\hbar$ the reduced Planck constant and $m$ the atomic mass of $^{87}$Rb). The dimensionless lattice depth $s$ is independently calibrated for each experiment~\cite{cabrera_2018}. We directly manipulate the lattice phase $\varphi(t)$ with $\varphi(0) = 0$ by varying the relative phase between the drives of the two acousto-optic modulators controlling the laser beams of the lattice~\cite{dupont_2021}. The hybrid trap potential $V_\text{hyb}$ has a small angular frequency $\omega_x = 2\pi \times 10$ Hz making it negligible at the timescales of the experiments presented here, which are driven by the lattice potential $V_\dL$.
In the subspace of null quasi-momentum, the external atomic state $\psi(x,t)$ is then represented by a superposition of plane waves:
\begin{equation}
\psi(x,t) = \sum_{\ell \in \mathbb{Z}} c_\ell(t) \chi_\ell(x),
\label{eq:psix}
\end{equation}
\noindent with $c_\ell(t) \in \mathbb{C}$, $\textstyle{\sum_\ell} |c_\ell(t)|^2 = 1$ and $\chi_\ell(x)=e^{i \ell k_\dL x}/\sqrt{d}$.

The experiment consists in continuously varying the reference position  of the lattice, given by $-\varphi(t)/k_\dL$, for $t\in \left[0, t_\dc\right]$ in order to control the final state $\psi(x,t_\dc)$ (see section \ref{subsec:qoc}).
All the traps are then suddenly switched off and the BEC goes into ballistic expansion. After a sufficiently long time-of-flight ($35\,\mathrm{ms}
$ for the data presented here), we measure the relative atomic populations in the different diffraction orders which correspond to the  probabilities $|c_\ell(t_\dc)|^2$. In order to completely reconstruct the quantum state $\psi(x,t_\dc)$, we also need to access the phases of the $c_\ell(t_\dc)$ coefficients. For this purpose, we sample with independent realizations the evolution of the prepared state in the static lattice (with $\varphi(t>t_\dc) = 0$) and use these data for a full state reconstruction (see section \ref{subseq:rec}).	

\subsection{Quantum optimal control}
\label{subsec:qoc}

\begin{figure}
	\begin{center}
		\includegraphics[scale=1]{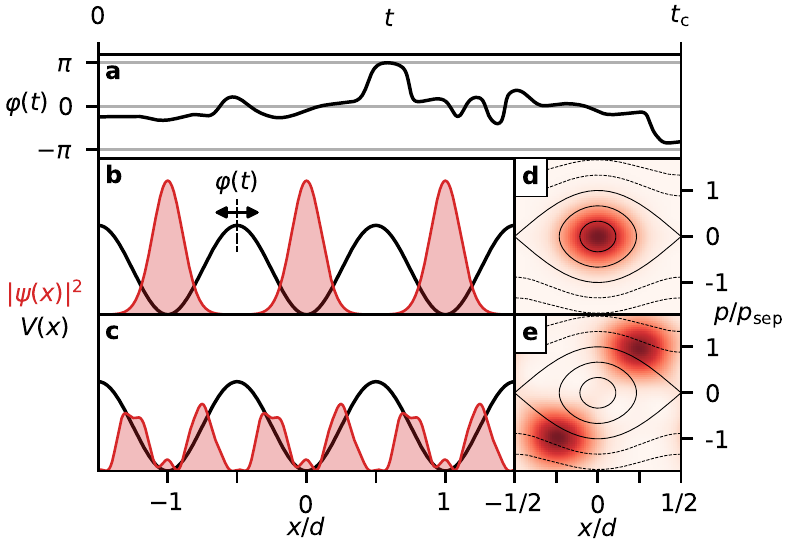}
		\caption{\textbf{Numerical demonstration of an optimized control field} for the preparation of a superposition of Gaussian states centered in $(x,p) = \pm(d/4, p_\dsep)$ (see section \ref{sec:unsqueezed_states}) starting from the ground state of the lattice with a numerical fidelity to target $\mathcal{F}_\dth>0.995$. \textbf{(a)} Time-evolution of the lattice phase $\varphi(t)$. \textbf{(b)} (resp. \textbf{(c)}) Squared modulus of the initial (resp. final) wave function in the $x$-representation (red) and lattice potential (black) over 3 lattice sites. \textbf{(d)} (resp. \textbf{(e)}) $(x,p)$ phase space, where the Husimi distribution corresponding to the states in \textbf{(b)} (resp. \textbf{(c)}) are depicted (red), as well as the classical trajectories in the static lattice at depth $s$ (black lines). The colorscale for each Husimi distribution extends from 0 to its maximum value. Parameters: $s = 5.55$ and $t_\dc = 1.75\,T_0 \approx 103.3$ $\upmu$s (see text).}
		\label{fig:figure1}
	\end{center}
\end{figure}

In order to reach a target state $\ket{\psi_\dT}$ from the ground state of the lattice potential defined in Eq.~\eqref{eq:global_potential}, we engineer the evolution of the control parameter $\varphi(t)$ over the duration $t_\dc$ using a first-order gradient-based optimal control algorithm~\cite{khaneja_2005,werschnik_2007,boscain_2021}. The algorithm consists in the iterative maximization of a figure of merit $\mathcal{F}$ that quantifies the success of applying a given $\varphi(t)$ to reach the target through integration of Schr\"odinger's equation. Our numerical method is detailed in~\cite{dupont_2021}. The figure of merit is the usual quantum fidelity $\mathcal{F}\left(\ket{\psi_\dT},\ket{\psi(t_\dc)}\right) = \left|\bra{\psi_\dT}\ket{\psi(t_\dc)}\right|^2$ and no constraint is put on $\varphi(t)$. The time $t_\dc$ is also fixed beforehand to $1.75\,T_0$ or $2\,T_0$ depending on the complexity of the preparation, with $T_0$ the period associated to the transition between the two lowest levels of the static lattice (\emph{e.g.} for $s = 5.5$, $T_0\approx59.3$ $\upmu$s). The state preparation is therefore clearly in the non-adiabatic regime. At the end of the optimization process, we obtain numerically a theoretical fidelity $\mathcal{F}_\dth = \left|\bra{\psi_\dT}\ket{\psi_\dQOC}\right|^2$ where $\ket{\psi_\dQOC}$ is the final state reached when using the optimized control field $\varphi_{\dQOC}(t)$. A typical result of a quantum optimal control query is shown in Fig.~\ref{fig:figure1}.

\subsection{Quantum state reconstruction}
\label{subseq:rec}

To ensure the quality of the quantum control scheme, we certify the preparation of the desired state by state tomography through likelihood maximization. Such a reconstruction of an experimentally prepared state requires finding the density matrix $\hatrho=\hatrho_\dML$ which maximizes the likelihood $\mathcal{L}$~\cite{lvovsky_2001,lvovsky_2004,deleglise_2008,Evrard2021,brown_2022}:
\begin{equation}
\hatrho_\dML = \argmax{{\mathcal{L}\left[\hatrho\right]}}
\quad \text{with} \quad
\mathcal{L}\left[ \hatrho \right] = \prod_j \pi_j^{f_j},
\label{eq:likelihood}
\end{equation}
\noindent where $\pi_j = \tr\lbrace\hatrho\hatE_j\rbrace$ are the expected measurement probabilities obtained from a set of operators $\hatE_j$ forming a positive operator-valued measure (POVM) and $f_j$ are the corresponding frequencies measured experimentally. In our case, the measurement frequencies $f_j$ are the relative populations of the plane waves $\ell$ measured at regularly spaced times $t \in \left[ t_\dc, t_\dc+t_\drec \right]$, divided by the number of sample times $N_t$ ($t_\drec = 100$ $\upmu$s and $N_t=20$ for data presented here unless specified otherwise):
\begin{equation}
	f_j = f_{\ell,t} = \frac{1}{N_t}|c_\ell(t)|^2
\end{equation}
As we intend to use these measurements to reconstruct the state prepared at $t_\dc$, the elements of the POVM are therefore:
\begin{equation}
\hatE_j = \hatE_{\ell,t} = \frac{1}{N_t}\hat{U}^{\dagger}(t,t_\dc)\ketbra{\chi_\ell} \hat{U}(t,t_\dc),
\end{equation}
\noindent with $\hatU(t,t_\dc)$ the evolution operator in the static lattice potential $V_\dL$ (with $\varphi=0$) from $t_\dc$ to $t$.

\begin{figure}
	\begin{center}
		\includegraphics[scale=1]{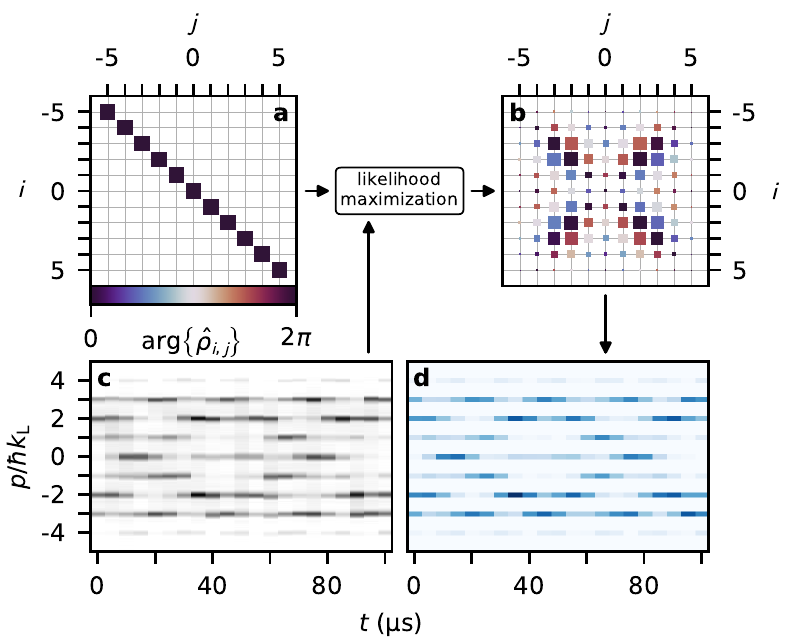}
		\caption{\textbf{Quantum state reconstruction by likelihood maximization} of the state experimentally prepared by the control field of Fig.~\ref{fig:figure1}. \textbf{(a-b)} Density matrices $\hatrho$ with $\text{arg}\left\lbrace \hatrho_{i,j} \right\rbrace$ color coded and $\left| \hatrho_{i,j} \right|$ size coded (not to scale between panels). \textbf{(a)} Identity $\mathbb{I}_D/D$ ($D=11$) as the initial guess. \textbf{(b)} Density matrix of maximum likelihood $\hatrho_\dML$. \textbf{(c)} Stack of experimental integrated absorption images taken during the evolution of the prepared state in the static lattice at $s = 5.5 \pm 0.5$. \textbf{(d)} Diagonal terms of the numerical propagation of $\hatrho_\dML$ which correspond to the absorption images of \textbf{(c)}.
		}
		\label{fig:figure2}
	\end{center}
\end{figure}

To obtain the maximum likelihood estimate of the 1-body density matrix $\hatrho_\dML$, we implement the iterative method developed in~\cite{rehacek_2001,lvovsky_2004}. We define a transformation $\hatrho^{(n)} \mapsto \hatrho^{(n+1)}$ such that $\mathcal{L} [\hatrho^{(n+1)}] \geq \mathcal{L}[\hatrho^{(n)}]$ and $\hatrho_\dML$ is a fixed point of the transformation. This algorithm reads:

\begin{itemize}
\item[1.] Set an initial guess state $\hatrho^{(0)}$,
\item[2.] Construct $\textstyle{R\left[\hatrho^{(0)}\right] = \sum_j f_j E_j/\tr{\hatrho^{(0)}E_j}}$,
\item[3.] Transform $\textstyle{\hatrho^{(0)} \mapsto \hatrho^{(1)} = R\left[\hatrho^{(0)}\right]\hatrho^{(0)}R\left[\hatrho^{(0)}\right]}$,
\item[4.] Repeat from step 2 until $\textstyle{\mathcal{L}\left[ \hatrho^{(n)}\right] - \mathcal{L}\left[ \hatrho^{(n-1)}\right]} \approx 0 \Leftrightarrow \hatrho^{(n)}$ converged to $\hatrho_\dML$.
\end{itemize}
Here, we use $\hatrho^{(0)} = \mathbb{I}_D/D$, with $D=2 \ell_\dmax+3$ chosen as the cut-off dimension of the Hilbert space so as to avoid boundary effects, $\ell_\dmax$ being the highest diffraction order at which some signal is experimentally detected  (for the experiments presented in this paper $2\leq\ell_\dmax\leq6$). Our choice of $\hatrho^{(0)}$ corresponds to the guess with the least initial information.
Finally, two indicators are computed to certify the preparation: the fidelity of $\hatrho_\dML$ to the numerically propagated state $\mathcal{F}_\dexp = \left<\psi_\dQOC|\hatrho_\dML|\psi_\dQOC\right>$ and the purity $\gamma = \tr{\hatrho^2_\dML}$ which is an indicator of our preparation reproducibility over the typically 20 realizations used for reconstruction. An accurate determination of the prepared state therefore requires a fine degree of reproducibility.
We illustrate the quantum state reconstruction process with an example in Fig.~\ref{fig:figure2}.

Even though interactions are present within the BEC, we verified through numerical simulations that their impact on the dynamics is negligible for evolution times lower than typically 150~$\upmu$s. This permits the use of the Schr\"odinger equation in both the optimal control and reconstruction algorithms.

\section{Non-squeezed Gaussian states}
\label{sec:unsqueezed_states}

In a first set of experiments, we prepare and reconstruct non-squeezed Gaussian states at arbitrary positions in phase space. In an analogous manner to the definition of coherent states in quantum optics \cite{Glauber1963,Bahr2007}, we define, at each lattice depth $s$, a non-squeezed Gaussian state $\ket{g(0,0)}$ as the ground state of the harmonic oscillator that approximates the bottom of each lattice well. For $s\gg 1$, the state $\ket{g(0,0)}$ can be equated to the ground state of the sinusoidal potential. We denote more generally as $\ket{g(u,v)}$ this same state displaced in phase space by $(u,v) = (k_\dL\left<\hatx\right>_{g(u,v)}, \left<\hatp\right>_{{g(u,v)}}/\hbar k_\dL)$. The displaced Gaussian state $\ket{g(u,v)}$ can be expanded on the plane wave basis with the coefficients \footnote{Eq.~\eqref{eq:unsqueezed_gaussian_states} only yields normalized states when the standard deviation of the envelope is far greater than the spacing of the momentum comb of the lattice, that is $\Delta p_0 \gg \hbar k_L \Leftrightarrow s \gg 1$, so we systematically renormalize our target states. For the non-squeezed states at depths $s \geq 5$, one can compute from Eq.~\eqref{eq:unsqueezed_gaussian_states}: ${\left|\sum\nolimits_{\ell\in\mathbb{Z}} |c_\ell(u,v)|^2 - 1\right| \leq 3.25\cdot10^{-5}}$.}:
\begin{equation}
c_\ell(u,v) = \left(\frac{2}{\pi \sqrt{s}}\right)^{1/4} e^{i uv/2} e^{-i \ell u} e^{-(\ell - v)^2/\sqrt{s}},
\label{eq:unsqueezed_gaussian_states}
\end{equation}
\noindent giving the position and momentum standard deviations in state $\ket{g}$: $k_\dL\Delta x_0  = s^{-1/4}$ and $\Delta p_0 /\hbar k_\dL= s^{1/4}/2$.
To relate our results to the classical phase space of the system, we also define $p_\dsep = \sqrt{s}\,\hbar k_\dL$, the positive momentum of the separatrix at $x=0$ (see \emph{e.g.} Fig.~\ref{fig:figure1}(d)).

\begin{figure}
	\begin{center}
		\includegraphics[scale=1]{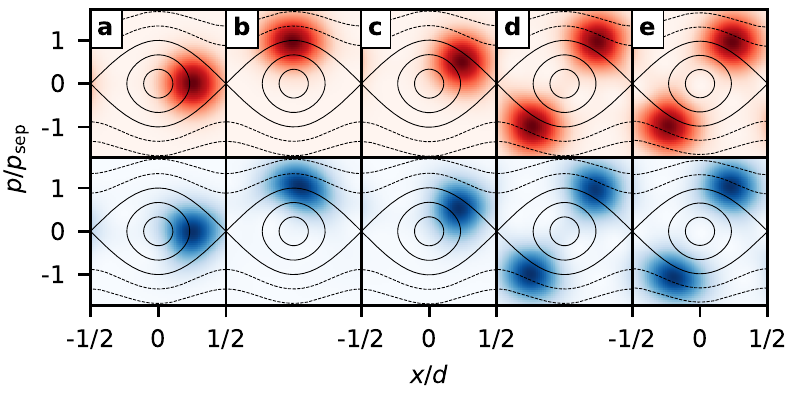}
		\caption{\textbf{Non-squeezed Gaussian states.} \textbf{(a-e)} Husimi representations in the phase space of the static lattice. Top (red): states $\ket{\psi_\dQOC}$ numerically prepared by optimal control. Bottom (blue): density matrices $\hatrho_\dML$ reconstructed from experimental data by likelihood maximization. The relative phases in the superpositions \textbf{(d-e)} are respectively 0 and $\pi$ (see text). The colorscale for each Husimi distribution extends from 0 to its maximum value. See Table~\ref{table1} for associated experimental parameters and figures of merit.}
		\label{fig:figure3}
	\end{center}
\end{figure}

In Fig.~\ref{fig:figure3}, we show the Husimi distributions $H(u,v)=\bra{g(u,v)}\hatrho\ket{g(u,v)}/2\pi$ of numerically propagated final states $\hat{\rho}_\dQOC=\ket{\psi_\dQOC}\bra{\psi_\dQOC}$ and of corresponding density matrices $\hatrho_\dML$ reconstructed from experimental data. The results are detailed in Table~\ref{table1}. We prepare translated non-squeezed Gaussian states with high fidelity to numerical simulations and good purity ($\gamma \geq 0.95$).
As our experimental reconstruction data come from several independent initial states evolved for different durations before measurement, we expect the decrease in purity $\gamma$ to result from residual experimental fluctuations.
In Fig.~\ref{fig:figure3}(d-e), we realize even and odd superpositions of non-squeezed Gaussian states, that is $\ket{\psi_T} = (\ket{g(u,v)} + e^{i\phi}\ket{g(-u,-v)})/\sqrt{2}$, with $\phi = 0,\pi$.
To our knowledge, there is no adiabatic method for preparing such superpositions of translated Gaussian or ground states in the lattice.
The differences between their momentum evolutions (see Appendix~\ref{appen:superposed_state}) allow to unambiguously identify that the prepared states are consistent with numerical simulations (Table~\ref{table1}), which is further confirmed by the very low cross fidelities: $\textstyle{\mathcal{F}(\psi_\dQOC^{(e)},\hatrho_\dML^{(d)}) = 0.004}$ and $\textstyle{\mathcal{F}(\psi_\dQOC^{(d)},\hatrho_\dML^{(e)}) = 0.008}$. The Husimi representations of the states (both for $\ket{\psi_\dQOC}$ and $\hatrho_\dML$) show however very little difference between the superposition states of opposite parity, a known feature of this quasi-distribution~\cite{HarocheRaimond06}.

\begin{table}
\begin{center}
\begin{tabular}{
>{\centering}p{0.175\linewidth}
>{\centering}p{0.145\linewidth}
>{\centering}p{0.145\linewidth}
>{\centering}p{0.145\linewidth}
>{\centering}p{0.145\linewidth}
>{\centering\arraybackslash}p{0.145\linewidth}
}
\hhline{======}
Fig.~\ref{fig:figure3} &  \textbf{a} & \textbf{b} & \textbf{c} & \textbf{d} & \textbf{e} \\ \hline
$u$ & $\pi/2$ & 0 & $\pi/2$ & $\pm\pi/2$ & $\pm\pi/2$ \\
$v$ & 0 & $\sqrt{s}$ & $\sqrt{s}/2$ & $\pm\sqrt{s}$ & $\pm\sqrt{s}$ \\
$\mathcal{F}_\dexp$ & 0.95 & 0.85 & 0.95 & 0.98 & 0.95 \\
$\gamma$ & 0.95 & 0.96 & 0.96 & 1.00 &  1.00 \\
$s$ & $5.50{\scriptstyle\pm0.25}$ & $5.49{\scriptstyle\pm0.20}$ & $5.57{\scriptstyle\pm0.20}$ & $5.5{\scriptstyle\pm0.5}$ & $5.30{\scriptstyle\pm0.25}$ \\
\hhline{======}
\end{tabular}
\caption{\textbf{Parameters used for the preparation of non-squeezed Gaussian states and figures of merit obtained from their reconstruction.} For all experiments $\mathcal{F}_\dth > 0.995$ and $t_\dc=1.75\,T_0$.}
\label{table1}
\end{center}
\end{table}

\section{Squeezed Gaussian states}
\label{sec:squeezing}

In a second set of experiments, we apply our preparation and reconstruction procedures to the squeezing of Gaussian states. We define the $x$-squeezing parameter as the ratio of standard deviations $\xi = \Delta x/\Delta x_0 = (\Delta p/\Delta p_0)^{-1}$
. Including $\xi$ in the definition of our Gaussian states, Eq.~\eqref{eq:unsqueezed_gaussian_states} becomes:
\begin{equation}
c_\ell^{(\xi)}(u,v) = \left(\frac{2\xi^2}{\pi \sqrt{s}}\right)^{1/4} e^{i uv/2} e^{-i \ell u} e^{-\xi^2(\ell - v)^2/\sqrt{s}}.
\label{eq:squeezed_state}
\end{equation}
\noindent For the squeezed Gaussian state $\ket{g^{(\xi)}}$ at lattice depth $s$, position and momentum standard deviations are given by $k_\dL\Delta x  = \xi\,s^{-1/4}$ and $\Delta p/\hbar k_\dL =s^{1/4}/2\,\xi$. The highest bound on $\xi$ is reached when only a single diffraction order is populated, which we can achieve up to $\abs{\ell}=10$~\cite{dupont_2021}. 

\begin{figure}
	\begin{center}
		\includegraphics[scale=1]{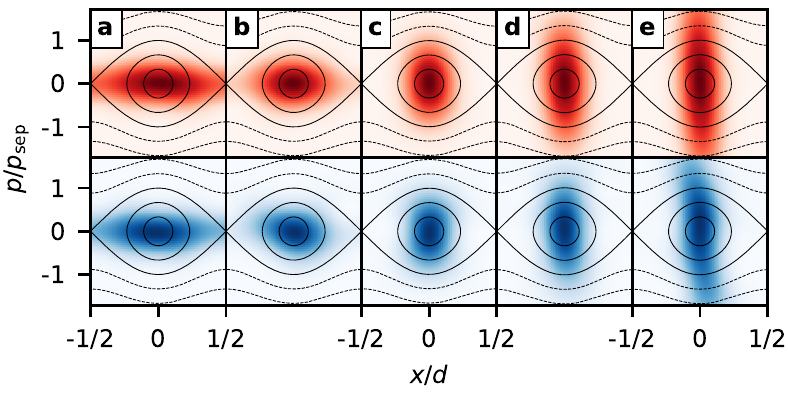}
		\caption{\textbf{Squeezed Gaussian states.} \textbf{(a-e)} Husimi representations in the phase space of the static lattice. Top (red): states $\ket{\psi_\dQOC}$ numerically prepared by optimal control. Bottom (blue): density matrices $\hatrho_\dML$ reconstructed from experimental data by likelihood maximization. The colorscale for each Husimi distribution extends from 0 to its maximum value.
		See Table~\ref{table2} for associated experimental parameters and figures of merit.}
		\label{fig:figure4}
	\end{center}
\end{figure}

Figure~\ref{fig:figure4} and Table~\ref{table2} display results for $(u,v)=(0,0)$ and $1/\xi$ ranging from $0.44$ to $4.34$. Up to $1/\xi = 2.75$, we prepare and reconstruct states with good fidelities and purities ($\mathcal{F}_\dexp\geq 0.92$ and $\gamma\geq 0.91$). For the highly squeezed state $1/\xi = 4.34$ of Fig.~\ref{fig:figure4}(e), it is necessary to increase $t_\dc$ to $2 \,T_0$ in order to attain a reasonable numerical fidelity $\mathcal{F}_\dth$. This is due to the complexity of the target state which consists in the superposition of 13 significantly populated momentum components ($|c_{|\ell|<7}|^2 > 0.025$) with as many complex coefficients to control. The simultaneous population of many momentum components has an even worse effect on the reconstruction as it significantly reduces the signal-to-noise ratio due to the lower number of atoms per diffraction peak (see Appendix~\ref{appen:squeezed_state}), which also requires an increase in reconstruction parameters $t_\drec$ and $N_t$ (to $125\,\upmu\mathrm{s}$ and $25$, respectively). Nevertheless, we achieve a fidelity $\mathcal{F}_\dexp>0.8$ even in that extreme case, and all the Husimi representations of Fig.~\ref{fig:figure4} show qualitatively very good agreement between $\hatrho_\dML$ and $\ket{\psi_\dQOC}$ for the squeezing of Gaussian states.

Interestingly, a squeezed state produced by Eq.~\eqref{eq:squeezed_state} at depth $s$ with squeezing parameter $\xi$ can be identified to a non-squeezed state produced by Eq.~\eqref{eq:unsqueezed_gaussian_states} at depth $s_\deff$. This leads to an effective lattice depth associated to the squeezed state $s_\deff = \xi^4 s$.
In that sense, Fig.~\ref{fig:figure4}(e) is the effective realization of the ground state of a lattice of depth $s_\deff \approx 2000$ in our lattice of depth $s = 5.62$. This is, to our knowledge, the first realization of such a state, the production of which is technically impossible with adiabatic methods. For example, with our setup, we would require a laser power of about 750~W in order to reach this lattice depth.

We also targeted Gaussian states both squeezed and rotated in the $(x,p)$ plane. Target state definition and results for those experiments are presented in Appendix~\ref{appen:rot_state}.

\begin{table}
\begin{center}
\begin{tabular}{
>{\centering}p{0.175\linewidth}
>{\centering}p{0.145\linewidth}
>{\centering}p{0.145\linewidth}
>{\centering}p{0.145\linewidth}
>{\centering}p{0.145\linewidth}
>{\centering\arraybackslash}p{0.145\linewidth}
}
\hhline{======}
Fig.~\ref{fig:figure4} &  \textbf{a} & \textbf{b} & \textbf{c} & \textbf{d} & \textbf{e} \\ \hline
$1/\xi$ & 0.44 & 0.62 & 1.65 & 2.75 & 4.34 \\
$\mathcal{F}_\dth$ & \multicolumn{3}{c}{\; \centeredhrulefill \; $> 0.995$ \; \centeredhrulefill \;} & 0.980 & 0.965 \\
$\mathcal{F}_\dexp$ & 0.99 & 0.96 & 0.97 & 0.92 & 0.81 \\
$\gamma$ & 1.00 & 1.00 & 0.99 & 0.91 & 0.83\\
$s$ & $5.49{\scriptstyle\pm0.20}$ & $5.49{\scriptstyle\pm0.20}$ & $5.45{\scriptstyle\pm0.40}$ & $5.57{\scriptstyle\pm0.20}$ & $5.62{\scriptstyle\pm0.25}$ \\
$t_\dc/T_0$ & \multicolumn{4}{c}{\; \centeredhrulefill \; 1.75 \; \centeredhrulefill \;} & 2 \\
$t_\drec$ ($\upmu$s) & \multicolumn{4}{c}{\; \centeredhrulefill \; 100 \; \centeredhrulefill \;} & 125 \\ \hhline{======}
\end{tabular}
\caption{\textbf{Parameters used for the preparation of squeezed Gaussian states and figures of merit obtained from their reconstruction.} For all experiments $(u,v) = (0,0)$.}
\label{table2}
\end{center}
\end{table}

\section{Enhancing a dynamical tunneling quantum simulation}
\label{sec:dynamical_tunneling}

\begin{figure}
	\begin{center}
		\includegraphics[scale=1]{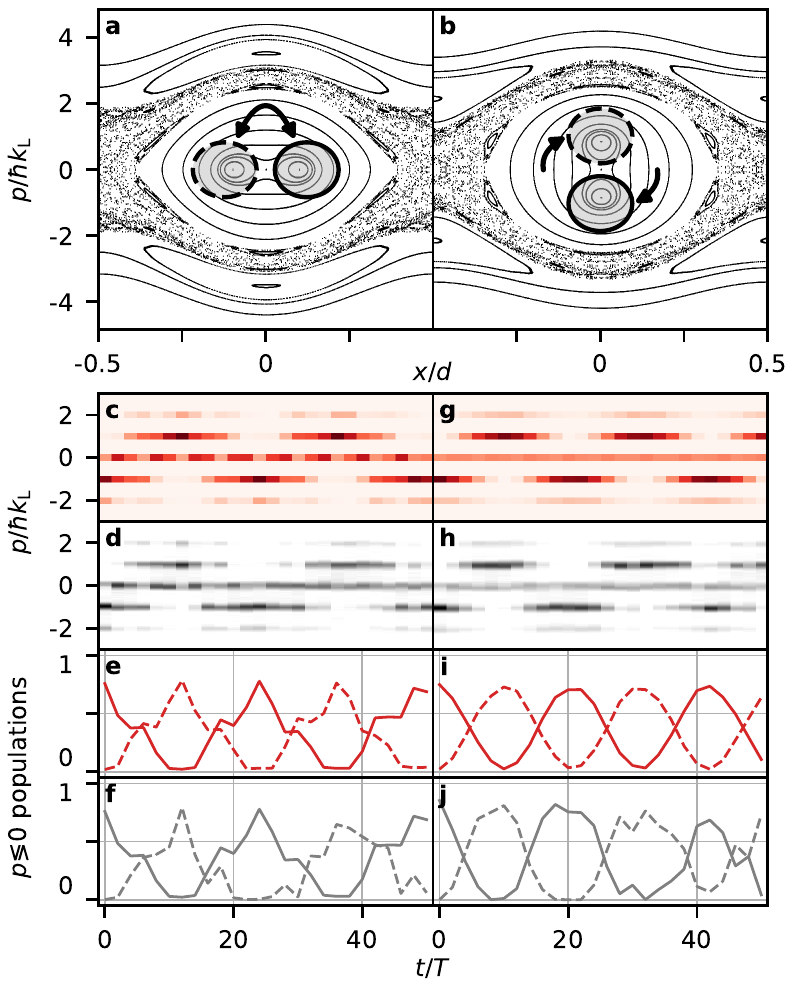}
		\caption{\textbf{Dynamical tunneling experiments.} \textbf{(a)} Poincar\'e section of the system with the initially populated regular region (gray area with black solid line border) and the coupled regular region that gets populated through dynamical tunneling (gray area with black dashed line border). \textbf{(b)} Same as \textbf{(a)} after phase space rotation (see text). \textbf{(c-f)} Results for the initially translated ground state. \textbf{(g-j)} Results with the initial state obtained from optimal control. \textbf{(c,g)} Numerical evolution of the momentum distribution. \textbf{(d,h)} Corresponding stack of experimental integrated absorption images. \textbf{(e,i)} Numerical and \textbf{(f,j)} experimental evolutions of the negative (solid line) and positive (dashed line) momentum populations. Parameters: \textbf{(c-f)} $s = 7.95 \pm 0.40$ and \textbf{(g-j)} $s = 7.95 \pm 0.30$.}
		\label{fig:figure5}
	\end{center}
\end{figure}

As a use-case example, we apply our QOC method to the production of the initial state for a quantum simulation experiment in a Floquet system. More precisely, we employ our protocol to prepare the optimal initial state for the observation of dynamical tunneling in an amplitude-modulated one-dimensional optical lattice.
In the mixed phase space of a periodically driven dynamical system, classical trajectories are either quasi-periodic (regular) or chaotic (resulting respectively in continuous lines or spread points in the Poincar\'e section, see \emph{e.g.} Fig.~\ref{fig:figure5}(a)). Quantum particles in such a system can undergo dynamical tunneling, oscillating from one region of regular trajectories to another, crossing classically impassable Kolmogorov-Arnold-Moser surfaces~\cite{book_keshavamurthy_2011}. For time-periodic Hamiltonians, a natural basis is the set of Floquet states, the eigenstates of the evolution operator over one period of modulation (the Floquet operator). Dynamical tunneling occurs when two non-degenerate Floquet states span the same regular regions of phase space, with a tunnelling oscillation frequency proportional to the quasi-energy difference between the two states in the Floquet spectrum~\cite{heller_1981, book_keshavamurthy_2011}.

In previous experiments with cold atoms in optical lattices, dynamical tunneling was studied with an initial sudden shift of the lattice to bring the ground state of the system in one of the tunnel-coupled regular regions~\cite{hensinger_2001, steck_2001, arnal_2020}.
Although this method provides evidence of the phenomenon, more than one frequency is observed in the tunneling signal as the initial states project only partially in the subspace of the two relevant Floquet states. Moreover the visibility of the oscillations is limited by the unequal-weight projection onto these states. We propose quantum optimal control as a way to optimize the initial state for the observation of dynamical tunneling.

The modulated potential is:
\begin{equation}
V(x,t) = -\frac{s}{2}E_\dL \left(1+\varepsilon \cos(\omega t) \right) \cos\left(k_\dL x\right),
\label{eq:amp_mod_pot}
\end{equation}
\noindent which generates the mixed phase portrait of Fig.~\ref{fig:figure5}(a) for the parameters $s(\omega) = 0.25\,(\hbar \omega / E_\dL)^2$ and $\varepsilon = 0.1$~\footnote{The phase portrait is strictly invariant in dimensionless coordinates $\tilde{x}\propto x/d$ and $\tilde{p}\propto{p/(md\omega)}$.}. We focus on the center of the Poincar\'e section, where a classical particle, stroboscopically observed every two periods of modulation, is bound to the lateral harmonic oscillator-like region it started in. For the quantum counterpart, $\omega$ sets an effective reduced Planck constant $\hbar_\deff = 2\,E_\dL/\hbar \omega$ that we fix at $0.36$ for the dynamical tunneling timescale to be compatible with the two-period stroboscopic sampling. Our optimal control target is the state that maximizes the visibility of the tunneling oscillation, that is the equal-weight superposition of the two main Floquet states in the central regular region of Fig.~\ref{fig:figure5}(a), with a relative phase such that the atoms start on the right side (see Appendix~\ref{appen:dyn_tun}). We can achieve a theoretical preparation fidelity $\mathcal{F}_\dth \geq0.995$ After evolution in the modulated potential, and before time-of-flight measurement, we modulate the potential during an additional half-period to perform a $\pi/2$-rotation around the center of the phase space and convert the population in the right (resp. left) regular region into experimentally accessible negative (resp. positive) momentum components~\cite{arnal_2020} (Fig.~\ref{fig:figure5}(a-b)).

Figure~\ref{fig:figure5} compares the results of dynamical tunneling experiments when the initial state is either approximated by a translation of the ground state (Fig.~\ref{fig:figure5}(c-f)) or targeted by our optimal control method (Fig.~\ref{fig:figure5}(g-j))~\footnote{As in~\cite{arnal_2020}, these experiments are performed with smaller BECs, of typically $5\cdot 10^4$ atoms in this work.}. The spectral content of the oscillations is clearly refined when the 2-Floquet state superposition is prepared, resulting in a greater signal-to-noise ratio for the measurement of the atomic tunneling.

\newpage

\section{Conclusion and prospects}

Quantum optimal control is a powerful tool for engineering the external state of ultracold atoms in an optical lattice. We use it here to manipulate the phase-space distribution of atoms in the unit cells of a one-dimensional optical lattice. We are able to arbitrarily position, squeeze and superpose Gaussian states, creating exotic phase-space distributions, as well as to target Floquet states. Using iterative state reconstruction inspired by quantum optics methods~\cite{lvovsky_2004}, we certify our control protocol, showing a good reproducibility in the preparation of the desired states (indicated by the purity of the reconstructed state) with great fidelities to numerical simulations. In the last section, we use our quantum optimal control protocol for the preparation of a specific superposition of Floquet states in a dynamical tunneling experiment.

These results demonstrate the promises optimal control holds for applications to quantum simulation and metrology. With the demonstrated ability of our method to produce highly non-stationary quantum states up to four times narrower in position than the ground state of the lattice, we achieve the preparation of states that are technologically inaccessible through the preparation of a ground state on a lattice setup. The short timescale for this preparation is also typically well below the duration that would be required for an adiabatic loading of such a ground state. The optimal control algorithm allows us to approach the minimum time for the preparation of exotic states in the lattice (which is still constrained by lattice dynamics~\cite{dupont_2021}).

This work, through the controlled generation of highly squeezed states, paves the way to the investigation of interaction effects in the dynamics, and therefore to the development of control protocols that include interactions~\cite{Adriazola2022}. It could also be generalized to systems of higher dimensionality, where stronger interaction regimes come into play. Ultimately, experiments exploiting optimal control at each stage (preparation, manipulation and measurement) for enhanced performance can be envisioned.
Thus optimal control may in the future allow to approach ultimate performances on a given setup.

\section*{Acknowledgments}
The authors thank Maxime Martinez and Clément Sayrin for helpful discussions. This study has been (partially) supported through the EUR grant NanoX No. ANR-17-EURE-0009 in the framework of the ``Programme d'Investissements d’Avenir" and research funding Grant No. ANR-17-CE30-0024. N.D. and F.A. acknowledge support from R\'egion Occitanie and Universit\'e Toulouse III-Paul Sabatier.

\appendix

\section{Superposed Gaussian states}
\label{appen:superposed_state}

\begin{figure}
	\begin{center}
		\includegraphics[scale=1]{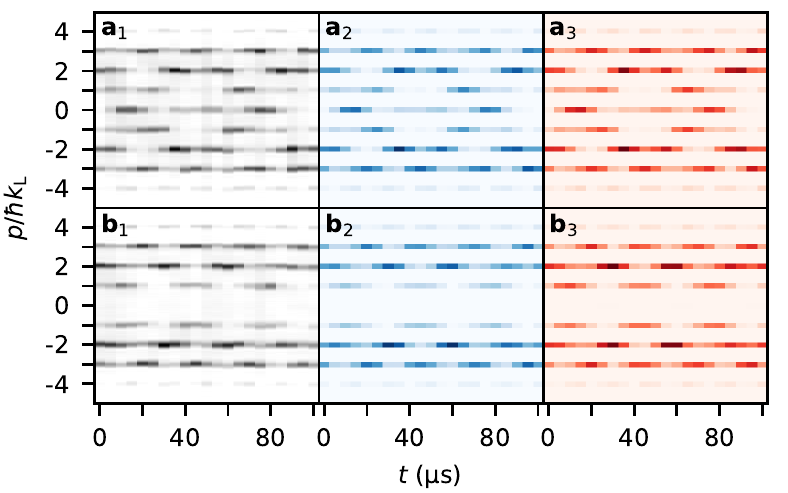}
		\caption{\textbf{Evolution of the momentum distribution of superpositions of Gaussian states kept in the static lattice}. Top panels \textbf{(a)} correspond to the even superposition of Fig.~\ref{fig:figure3}(d). Bottom panels \textbf{(b)} correspond to the odd superposition of Fig.~\ref{fig:figure3}(e). \textbf{(a$_1$,b$_1$)} Stacks of experimental integrated absorption images taken during the evolution of the prepared states in the static lattice. \textbf{(a$_2$,b$_2$)} Numerical evolution of the diagonal terms of the density matrices reconstructed from \textbf{(a$_1$,b$_1$)}. \textbf{(a$_3$,b$_3$)} Numerical evolution of the states $\ket{\psi_\dQOC}$ obtained by optimal control.  See Table~\ref{table1} for associated experimental parameters and figures of merit.}
		\label{fig:figure6}
	\end{center}
\end{figure}

We show on Fig.~\ref{fig:figure6}(a$_1$-b$_1$) the data used for the reconstruction of the even and odd superpositions of non-squeezed Gaussian states of Fig.~\ref{fig:figure3}(d-e). A striking difference between the evolution of their momentum distributions can be seen on the 0\textsuperscript{th} order of diffraction that turns on and off for the even superposition of Fig.~\ref{fig:figure6}(a) whereas it is rigorously off for the odd superposition of Fig.~\ref{fig:figure6}(b). Despite their almost indistinguishable Husimi distributions, the different time-evolutions of the two superpositions allow for the reconstruction of states that evolve very much like the experimental data~(Fig.~\ref{fig:figure6}(a$_2$-b$_2$)) as well as the numerically prepared states $\ket{\psi_\dQOC}$~(Fig.~\ref{fig:figure6}(a$_3$-b$_3$)).

\section{Highly squeezed Gaussian state}
\label{appen:squeezed_state}

\begin{figure}
	\begin{center}
		\includegraphics[scale=1]{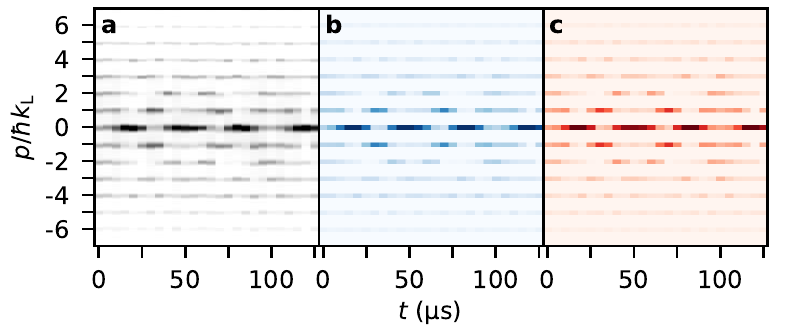}
		\caption{\textbf{Evolution of the momentum distribution of a highly squeezed state kept in the static lattice}. \textbf{(a)} Stack of experimental integrated absorption images taken during the evolution of the prepared state in the static lattice. \textbf{(b)} Numerical evolution of the diagonal terms of the density matrix reconstructed from \textbf{(a)}. \textbf{(c)} Numerical evolution of the state $\ket{\psi_\dQOC}$ obtained by optimal control. \textbf{(b)} and \textbf{(c)} correspond to the evolutions of the states presented in Fig.~\ref{fig:figure4}(e).}
		\label{fig:figure7}
	\end{center}
\end{figure}

In Figure~\ref{fig:figure7}(a), we show the reconstruction data for the highly squeezed state (1/$\xi = 4.34$) of Fig.~\ref{fig:figure4}(e). For $|\ell|>3$ we can see that the signal-to-noise ratio gets quite low, which, in addition to the higher number of plane wave coefficients to determine, greatly complicates the reconstruction process. Despite the lower fidelity $\mathcal{F}_\dexp=0.81$ of $\hatrho_\dML$ to $\ket{\psi_\dQOC}$ that we obtain, hardly any difference is visible between the numerical evolution of the momentum distributions associated with the reconstructed state (Fig.~\ref{fig:figure7}(b)) and with the prepared state (Fig.~\ref{fig:figure7}(c)).
This seems to indicate that the impact on the fidelity originates from differences in the extreme plane wave coefficients for which the signal-to-noise is lower. This sensitivity to noise also makes the data less reproducible, leading to a reduced state purity $\gamma = 0.83$.%

\section{Rotated squeezed Gaussian states}
\label{appen:rot_state}

\begin{figure}
	\begin{center}
		\includegraphics[scale=1]{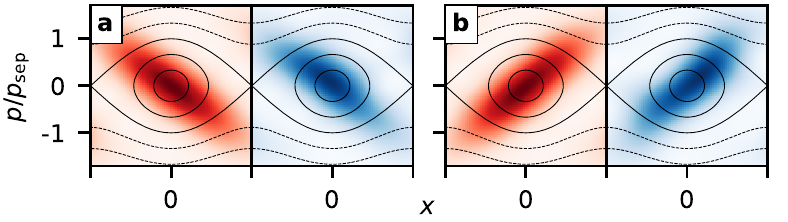}
		\caption{\textbf{Rotated squeezed Gaussian states}. \textbf{(a-b)} Husimi representations in the phase space of the static lattice. Left (red): state $\ket{\psi_\dQOC}$ numerically prepared by optimal control. Right (blue): density matrix $\hatrho_\dML$ reconstructed from experimental data by likelihood maximization. The colorscale for each Husimi distribution extends from 0 to its maximum value. See Table~\ref{table3} for associated experimental parameters and figures of merit.}
		\label{fig:figure8}
	\end{center}
\end{figure}

In order to perform the rotation in phase space of squeezed Gaussian states we need to redefine our target states. For a positive rotation angle $\theta$ in the $(x-p)$ phase space, the plane wave coefficients of these states are:
\begin{equation}
	c_\ell^{(\xi,\theta)}(u,v) = \left(\frac{\Re[A]}{\pi}\right)^{1/4} e^{i uv/2} e^{-ilu} e^{-A(l-v)^2/2},
	\label{eq:rotated_squeezed_state}
\end{equation}
\noindent with
\begin{equation*}
A = \frac{\cosh(r)-\sinh(r) \, e^{2i\theta}}{\cosh(r)+\sinh(r) \, e^{2i\theta}}
\quad \text{and} \quad
r = \frac{1}{4}\ln(\frac{s}{4\, \xi^4}).
\end{equation*}
Figure~\ref{fig:figure8} and Table~\ref{table3} show results for $(u,v) = (0,0)$, $\xi = 1/3$ and $\theta = \pm \pi/4$.

\begin{table}
\begin{center}
\begin{tabular}{
>{\centering}p{0.33\linewidth}
>{\centering}p{0.3\linewidth}
>{\centering\arraybackslash}p{0.3\linewidth}
}
\hhline{===}
Fig.~\ref{fig:figure8} &  \textbf{a} & \textbf{b} \\ \hline
$\theta$ & $\pi/4$ & $-\pi/4$ \\
$\mathcal{F}_\dexp$ & 0.91 & 0.89 \\
$\gamma$ & 0.88 & 0.90 \\
\hhline{===}

\end{tabular}
\caption{\textbf{Parameters used for the preparation of rotated squeezed Gaussian states and figures of merit obtained from their reconstruction.} For all experiments $(u,v) = (0,0)$, $\xi = 1/3$, $\mathcal{F}_\dth \geq 0.995$, $s = 5.45 \pm 0.30$ and $t_\dc = 1.75\,T_0$.}
\label{table3}
\end{center}
\end{table}

\section{Initial states for dynamical tunneling}
\label{appen:dyn_tun}

We define the ideal state for dynamical tunneling as the equal-weight superposition of the two Floquet states $\ket{F_\dA}$ and $\ket{F_\dB}$ that support the tunneling:
\begin{equation}
\ket{\psi(\theta)} = \frac{1}{\sqrt{2}}\left(\ket{F_\dA} + e^{i\theta} \ket{F_\dB}\right),
\label{eq:target_dynamical_tunneling}
\end{equation}
\noindent where $\ket{F_\dA}$ and $\ket{F_\dB}$ can be identified by their overlap with a non-squeezed Gaussian state  (Eq.~\eqref{eq:unsqueezed_gaussian_states}) centered in either lateral regular regions of the Poincar\'e section (Fig.~\ref{fig:figure5}(a)). With optimal control, we initialize the system in the right regular region, that is targeting the superposition $\ket{\psi(\theta_\dR)}$ with the phase difference between the two Floquet states:
\begin{equation}
\theta_\dR = \argmax{\left< \hat{x} \right>_{\psi(\theta)}}.
\label{eq:relative_phase_dynamical_tunneling}
\end{equation}
Defining $\ket{\phi_0}$, the ground state of the lattice, and $\hat{D}(\Delta x)$, the translation operator that translates in $x$ by a quantity $\Delta x$, the translated ground state to which we compare the preparation of $\ket{\psi(\theta_\dR)}$ is $\hat{D}(\Delta x_\dR)\ket{\phi_0}$ with $\Delta x_\dR$ maximizing the overlap between $\ket{\psi(\theta_\dR)}$ and the translated ground state:
\begin{equation}
\Delta x_\dR = \argmax{\left|\bra{\psi(\theta_\dR)}\hat{D}(\Delta x)\ket{\phi_0}\right|^2}.
\label{eq:delta_x_dynamical_tunneling}
\end{equation}
\noindent For our parameters, we find a fidelity between the translated ground state and the optimal state $\left|\bra{\psi(\theta_\dR)}\hat{D}(\Delta x_\dR)\ket{\phi_0}\right|^2 \approx 0.91$.

\begin{figure}
	\begin{center}
		\includegraphics[scale=0.95]{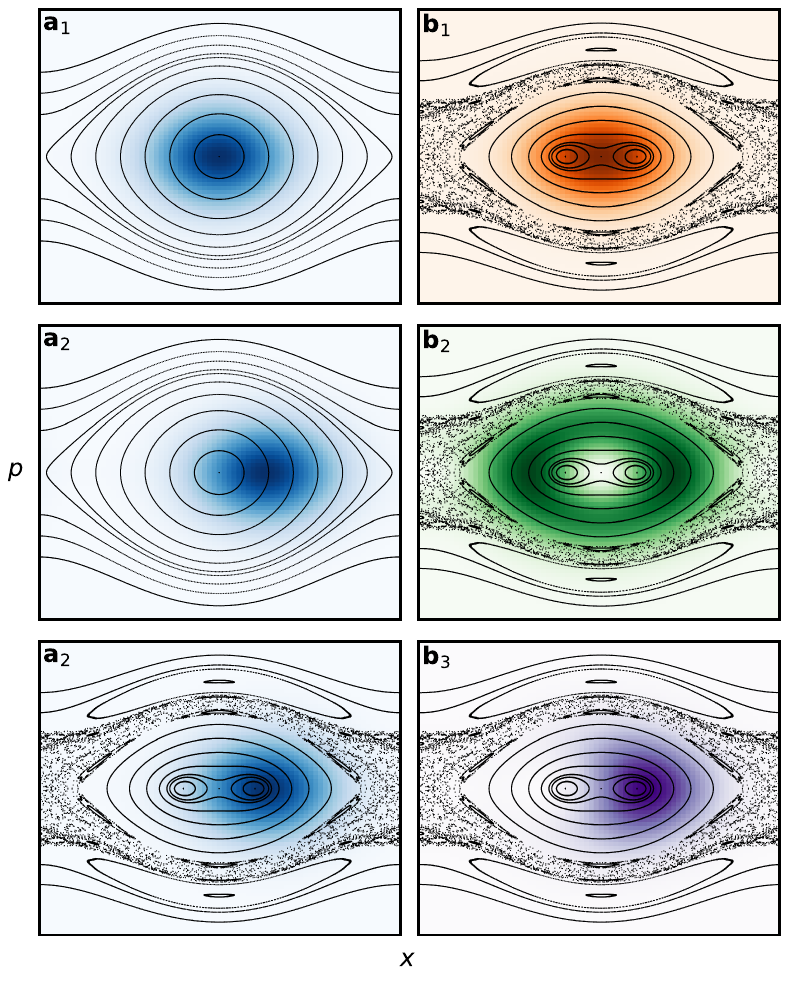}
		\caption{\textbf{Initial states for regular dynamical tunneling.} Husimi representations and underlying classical phase spaces for \textbf{(a)} the translation method and \textbf{(b)} the relevant Floquet states. \textbf{(a$_1$)} $\ket{\phi_0}$. \textbf{(a$_2$)} (resp. \textbf{(a$_3$)}) $\hat{D}(\Delta x_\dR)\ket{\phi_0}$ just before (resp. after) the start of the modulation. \textbf{(b$_1$)} $\ket{F_\dA}$. \textbf{(b$_2$)} $\ket{F_\dB}$. \textbf{(b$_3$)} $\ket{\psi(\theta_\dR)}$. See text for the definitions of these states. The colorscale for each Husimi distribution extends from 0 to its maximum value.}
		\label{fig:figure9}
	\end{center}
\end{figure}

Fig.~\ref{fig:figure9} shows the Husimi representations of $\ket{\phi_0}$, $\hat{D}(\Delta x_\dR)\ket{\phi_0}$, $\ket{F_\dA}$, $\ket{F_\dB}$ and $\ket{\psi(\theta_\dR)}$. The differences between the translated ground state $\hat{D}(\Delta x_\dR)\ket{\phi_0}$ and the optimal state $\ket{\psi(\theta_\dR)}$ may seem small, however their impact on the tunneling signal is quite important, leading to much sharper oscillations with the optimal state (see Fig.~\ref{fig:figure5}).

\bibliography{article_phase_space_bib}

\end{document}